\documentclass[letterpaper, 10 pt, conference]{ieeeconf}
\IEEEoverridecommandlockouts
\pdfoutput=1
\usepackage{amsmath}
\usepackage{amssymb}
\usepackage{color,soul}
\usepackage[pdftex]{graphicx}
\allowdisplaybreaks[4]
\usepackage[noadjust]{cite}
\usepackage{bm}

\usepackage{float}
\usepackage{comment}
\usepackage{epstopdf}
\usepackage{pgfplots, pgfplotstable}
  \pgfplotsset{compat=newest}
  \usetikzlibrary{plotmarks}
  \usetikzlibrary{arrows.meta}
  \usetikzlibrary{positioning,shapes,shadows,arrows,shapes.multipart}
  \usepgfplotslibrary{fillbetween}
  \usepgfplotslibrary{patchplots}
  \usepgfplotslibrary{groupplots}
  \usepackage{grffile}
  \usepackage{url}
\usepackage[nolist,printonlyused]{acronym}
\usepackage{listings}
\newfloat{lstfloat}{t}{lop}
\floatname{lstfloat}{Code Block}

\usepackage{algorithm}
\usepackage[algo2e, ruled]{algorithm2e}
\usepackage{xurl}
\usepackage{xfrac}
\usepackage{bbm}
\usepackage{todonotes}
\usepackage[font=small,skip=3pt]{caption}
\usetikzlibrary{external}
\begin{document}

\title{Adam-based Augmented Random Search for Control Policies for Distributed Energy Resource Cyber Attack Mitigation}

\author{\authorblockN{Daniel Arnold$^{\dagger, \star}$, Sy-Toan Ngo$^{\dagger, \star}$, Ciaran Roberts$^\dagger$, Yize Chen$^\dagger$, Anna Scaglione$^{\circ}$, Sean Peisert$^\dagger$}
\authorblockA{$^\dagger$\textit{Lawrence Berkeley National Laboratory}
\{dbarnold,sytoanngo,cmroberts,ychen,sppeisert\}@lbl.gov \\
$^\circ$\textit{Cornell Tech}
as337@cornell.edu \\
$^\star$\textit{These authors contributed equally to this effort}
} \\
}

% \author{\authorblockN{Michael Shell}\authorblockA{School of Electrical and\\Computer Engineering\\Georgia Institute of Technology\\Atlanta, Georgia 30332--0250\\Email: mshell@ece.gatech.edu}\and\authorblockN{Homer Simpson}\authorblockA{Twentieth Century Fox\\Springfield, USA\\Email: homer@thesimpsons.com}\and\authorblockN{James Kirk\\and Montgomery Scott}\authorblockA{Starfleet Academy\\San Francisco, California 96678-2391\\Telephone: (800) 555--1212\\Fax: (888) 555--1212}}

\maketitle

%%%%%%%%%%%%%%%%%%%%%%%%%%%%%%%%%%%%%%%%%%%%%%%%%%
%% SPONSOR FOOTNOTE
%%%%%%%%%%%%%%%%%%%%%%%%%%%%%%%%%%%%%%%%%%%%%%%%%%
\let\thefootnote\relax\footnote{This research was supported in part by the Director, Cybersecurity, Energy Security, and Emergency Response, Cybersecurity for Energy Delivery Systems program, of the U.S. Department of Energy, under contract DE-AC02-05CH11231.  Any opinions, findings, conclusions, or recommendations expressed in this material are those of the authors and do not necessarily reflect those of the sponsors of this work.}

\vspace{-0.3in}

\begin{abstract}
Volt-VAR and Volt-Watt control functions are mechanisms that are included in distributed energy resource (DER) power electronic inverters to mitigate excessively high or low voltages in distribution systems.  In the event that a subset of DER have had their Volt-VAR and Volt-Watt settings compromised as part of a cyber-attack, we propose a mechanism to control the remaining set of non-compromised DER to ameliorate large oscillations in system voltages and large voltage imbalances in real time.  To do so, we construct control policies for individual non-compromised DER, directly searching the policy space using an Adam-based augmented random search (ARS).  In this paper we show that, compared to previous efforts aimed at training policies for DER cybersecurity using deep reinforcement learning (DRL), the proposed approach is able to learn optimal (and sometimes linear) policies an order of magnitude faster than conventional DRL techniques (e.g., Proximal Policy Optimization).
\end{abstract}

% \begin{IEEEkeywords}
% Cyber Security, Deep Reinforcement Learning, Distributed Energy Resources, Smart Inverter
% \end{IEEEkeywords}

\section{Introduction}
\label{sec:intro}

Distributed Energy Resources (DER), most notably rooftop solar photovoltaic (PV) systems, are envisioned to be key components of realizing local, state, and federal renewable energy goals in the next several decades.  On the federal level, discussions of a national clean energy standard \cite{npr_clean_energy} and efforts to further bring down the cost of solar \cite{doe_cut_solar} will undoubtedly serve to accelerate already high levels of adoption of rooftop solar PV systems.  Faced with a future grid where a large amount of power is generated from DER, stakeholders have convened over the past two decades to codify standards that seek to govern the behavior of DER to ensure resiliency and reliability of the power system.

A standard that has garnered significant attention in recent years is IEEE 1547 \cite{IEEE_1547}.  
Among other things, this standard proposes mechanisms that allow DER smart inverters to adjust their active and reactive power outputs in response to changes in local system voltages.  In so doing, the standard seeks to ensure that DER will collectively act to minimize occurrences of over- or under-voltages.

Two mechanisms proposed within IEEE 1547 aimed at providing inverter-based voltage regulation are known as Volt-VAR (VV) and Volt-Watt (VW) control functions.  These control functions, depicted in Figs. \ref{fig:vvc} - \ref{fig:vwc}, alter DER reactive and active power injection according to piecewise linear non-increasing functions of the locally sensed voltage.  In addition to prescribing the general structure of the control functions, IEEE 1547 also specifies different parameterizations of the piecewise linear curves, which are realized via adjusting the parameters $\bm{\eta} = [\eta_{1},\eta_{2},\eta_{3},\eta_{4},\eta_{5}]$.  The ability to adjust the parameters of the VV/VW curves, in theory, allows the dynamic response of DER to changing system voltages to be tailored to specific networks, or regions within a given network.  

While the flexibility offered by IEEE 1547 ensures that DER dynamic voltage response can be adjusted to better serve different types of networks (e.g., radial vs. meshed systems), the ability to adjust DER dynamics coupled with Internet, cellular, or power line carrier connectivity of the inverters themselves, potentially allows remote updating of VV/VW parameters by an entity with malicious intention \cite{sahoo2019cyber}.  An excellent example of the extent to which aggregations of smart inverters can be remotely updated was illustrated in Hawaii, where local utilities worked with a smart inverter vendor to remotely update the autonomous control functions of 800,000 inverters in a single day \cite{spectrum2015}.  

In the event that a portion of the DER smart inverter functions in a given network have had their settings adjusted as part of a cyber-attack, our previous works (Roberts et. al. \cite{roberts2020deep,roberts2021deep_unbalance}) explored the use of Deep Reinforcement Learning (DRL) to determine optimal control policies that alter the behavior of the remaining population of \emph{non-compromised} DER to attempt to mitigate the effect of the cyber-attack in real time.  In our first paper, we utilized Proximal Policy Optimization (PPO) to train optimal polices that adjust DER smart inverter VV and VW functions to mitigate attacks aimed at creating large oscillations in system voltages \cite{roberts2020deep}.  In a subsequent paper, we extended this framework using a different reward function to develop optimal policies for mitigating attacks designed to create large voltage imbalances in multi-phase systems \cite{roberts2021deep_unbalance}.  

The PPO-based policies developed in our previous efforts proved successful in mitigating large oscillations and voltage imbalances even when almost half of the PV smart inverter control functions in a given network were compromised during the cyber-attack.  The main drawback of the DRL architecture used to train policies to manage non-compromised DER was lengthy training time.  Training on relatively small networks, such as the IEEE 37-node test feeder using modest resources (Intel® Core™ i7-8850H CPU @ 2.60GHz, 16GB RAM) took several hours.  Training control policies on the IEEE 8500 node test feeder would often take tens of hours.  While proper parallelization and HPC/cloud resources could be leveraged to bring down the overall training time and enable these solutions to scale to larger networks, many utilities (rural electric cooperatives in the U.S., for example) lack the expertise and financial resources to properly engage these assets for training unique policies for different networks.  In order to democratize the training of network-specific control policies that optimally manage DER to ensure grid stability in the face of cyber attacks, additional effort is needed to develop alternative optimal control strategies which can train equivalent policies in substantially less time.

In contrast to many popular reinforcement learning techniques, evolutionary strategies (ES) do not attempt to approximate both the value function and the optimal policy of a Markov decision process (MDP).  ES, rather, directly search for optimal policies through perturbing the present policy and evaluating the improvement in the reward by conducting ``rollouts" of the MDP under the perturbed policy.  Variants of this ``perturb and observe" paradigm include genetic algorithms (GAs) from stochastic optimization \cite{spall2003stochastic} as well as extremum seeking techniques from nonlinear control theory \cite{krstic2003extremum}.  In a compelling piece, Salimans et. al. \cite{salimans2017evolution} demonstrated that random search (an extremely simple type of evolutionary strategy) can be leveraged to learn competitive and highly scalable policies compared to popular RL techniques for several MDPs.  Extending this work, Mania et. al. \cite{mania2018simple} proposed the \emph{augmented random search} (ARS) algorithm that was used to train \emph{linear} policies for several challenging MDP learning tasks with excellent sample efficiency.

In this paper, motivated by the success of ARS in learning optimal policies for complex MDPs with improved sample efficiency, we consider an extension of ARS with an adaptive learning rate applied to learn optimal policies for DER cyber-resiliency.  By incorporating the \emph{Adam optimization method} \cite{duchi2011adaptive} into the gradient update step of ARS, we are able to train \emph{linear} policies to mitigate voltage oscillations and neural-network based policies to mitigate voltage imbalances caused by cyber-attacked DER.  Our experiments show that the Adam-based ARS approach (henceforth referred to as Adam-ARS) enables training of competitive policies an order of magnitude faster than PPO and with less variance than ARS.  As such, this work stands to enable efficient and scalable learning of optimal policies for controlling DER smart inverters on a network-specific basis.

This paper is organized as follows.  Dynamic models of the DER smart inverter VV and VW control functions, as well as descriptions of the Adam solver and ARS are discussed in Section \ref{sec:preliminaries}.  Section \ref{sec:methodology} presents the Adam-ARS algorithm and discusses the framework for training optimal policies.  Results showing the effectiveness of the trained policies in mitigating voltage oscillations and imbalances introduced by cyber-attacked DER are provided in Section \ref{sec:results}.  Finally, concluding remarks are provided in Section \ref{sec:conclusions}.

\section{Preliminaries}
\label{sec:preliminaries}

%%%%%%%%%%%%%%%%%%%%%%%%%%%%%%%%%%%%%%%%%%%%
\subsection{Inverter Dynamic Modeling and Agent Interaction}

\begin{figure}[h]
\includegraphics[width=0.48\textwidth]{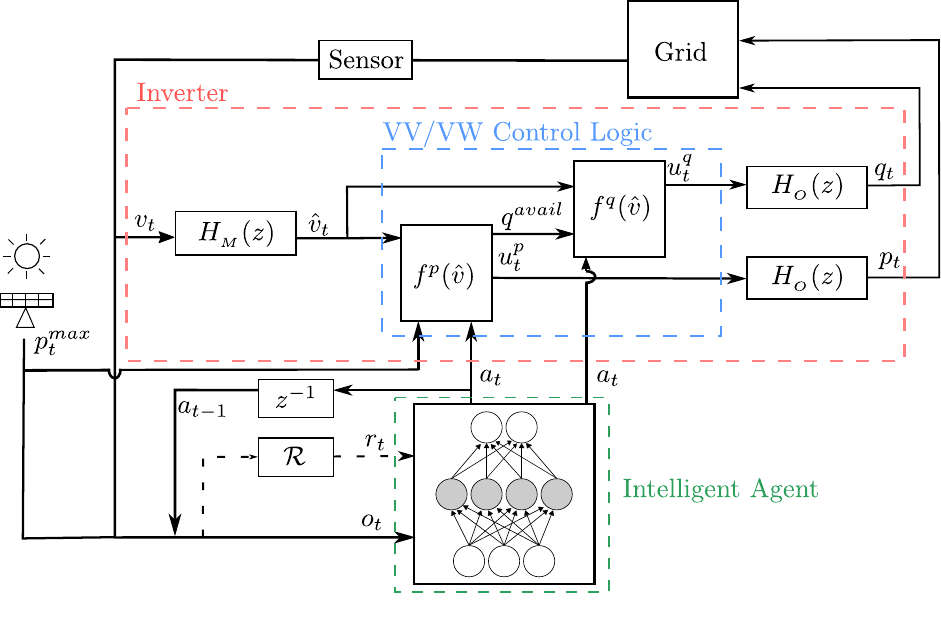}
\caption{Block diagram of VV and VW control logic of an inverter.}
\label{fig:inverter_block_diagram}
\end{figure}

The goal of this work is to train an intelligent agent capable of adjusting the parameters of VV/VW control functions in \emph{non-compromised} DER to mitigate large voltage oscillations and imbalances introduced by subsets of DER which have been cyber-attacked.  A depiction of the agent interacting with a dynamic model of a DER smart inverter is depicted in Fig. \ref{fig:inverter_block_diagram}.  

\begin{figure}[ht!]
\centering
\resizebox {21pc} {!} { \begin{tikzpicture}

% horizontal axis
\coordinate (xmax) at (5,0);
\coordinate (xmin) at (-5,0);
\draw[<->] (xmin) -- (xmax) node[anchor=west] {$V$};
    
% vertical axis
\coordinate (ymax) at (0,3);
\coordinate (ymin) at (0,-3);
\draw[<->] (ymin) -- (ymax) node[anchor=south] {\% available VARs};
    
% labels
\filldraw[black]	
		(-2,0) circle (2pt) node[anchor=north] {$\eta_{2}$}
		(2,0) circle (2pt) node[anchor=south] {$\eta_{3}$}
        
        (-4,2) circle (2pt)
        (4,-2) circle (2pt);
        
\draw	
		(0,2) node[anchor=west] {100\%}
		(-4,0) node[anchor=north] {$\eta_{1}$}
		(0,-2) node[anchor=east] {-100\%}
		(4,0) node[anchor=south] {$\eta_{4}$};

\draw[dotted] (4,0) -- (4,-2);
\draw[dotted] (-4,0) -- (-4,2);

\draw[dotted] (0,-2) -- (4,-2);
\draw[dotted] (0,2) -- (-4,2);

% curve
\draw[thick][<->] (-5,2) -- (-4,2) -- (-2,0) -- (2,0) -- (4,-2) -- (5,-2);

\end{tikzpicture} }
\caption{Inverter Volt-VAR curve.  Positive values denote VAR injection.}
\label{fig:vvc}
\end{figure}
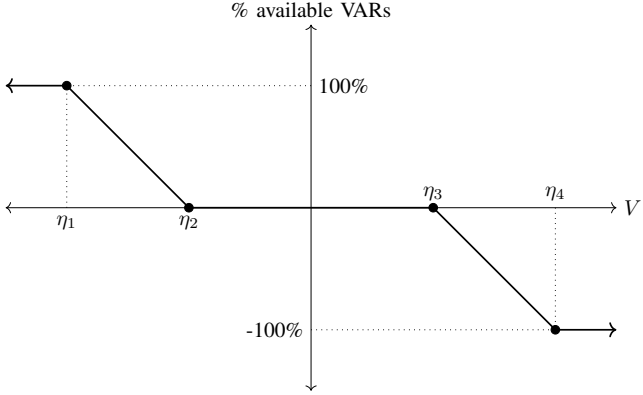

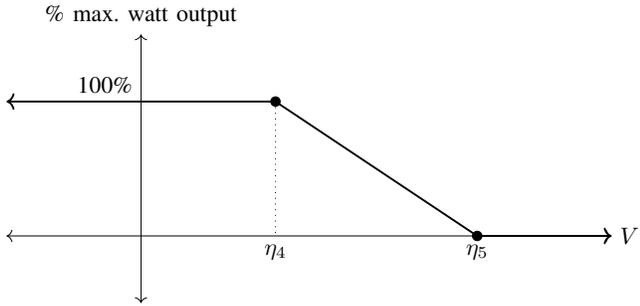
\begin{figure}[ht!]
\centering
\resizebox {21pc} {!} { \begin{tikzpicture}

% horizontal axis
\coordinate (xmax2) at (7,0);
\coordinate (xmin2) at (-2,0);
\draw[<->] (xmin2) -- (xmax2) node[anchor=west] {$V$};
    
% vertical axis
\coordinate (ymax2) at (0,3);
\coordinate (ymin2) at (0,-1);
\draw[<->] (ymin2) -- (ymax2) node[anchor=south] {\% max. watt output};
    
% labels
\draw
		%bottom plot
		%(1,0) node[anchor=north] {$\Delta V_{1}$}
        (2,0) node[anchor=north] {$\eta_{4}$}
        (0,2.25) node[anchor=east] {100\%};

\filldraw[black]
		%bottom plot
        %(1,2) circle (2pt)
        (5,0) circle (2pt) node[anchor=north] {$\eta_{5}$}
        (2,2) circle (2pt);

% % ranges
\draw[dotted] (2,2) -- (2,0);
%\draw[dotted] (1,2) -- (1,0);

% curve
\draw[thick][<->] (-2,2) -- (2,2) -- (5,0) -- (7,0);

\end{tikzpicture} }
\caption{Inverter Volt-Watt curve.  Positive values denote watt injection.}
\label{fig:vwc}
\end{figure}

As is shown in the figure, the smart inverter dynamic model maps grid voltages $v_{t}$ into active $p_{t}$ and reactive $q_{t}$ power injections.  The grid voltage is first passed through a low-pass filter $H_{M}(z)$ to generate the measured grid voltage $\hat{v}_{t}$ which is fed into the nonlinear Volt-VAR and Volt-Watt control functions, represented by $f^{q}(\hat{v})$ and $f^{p}(\hat{v})$.  These functions compute reactive and active power setpoints $u^{q}_{t}$ and $u^{p}_{t}$ which are passed through low pass filters $H_{o}(z)$ to create the actual powers injected into the grid, $q_{t}$ and $p_{t}$. The VW/VV control functions are shown in Figs. \ref{fig:vvc} - \ref{fig:vwc} and consist of piecewise linear non-increasing functions of grid voltage parameterized by a set of voltage values $\bm{\eta} = [\eta_{1}, \eta_{2}, \eta_{3}, \eta_{4}, \eta_{5}]$.  The VV/VW control structure depicted in Fig. \ref{fig:inverter_block_diagram} is known as Volt-Watt precedence \cite{inverter2016} where priority is given to the VW controller to compute the active power setpoint $u^{p}_{t}$.  Following the computation of $u^{p}_{t}$, any additional inverter capacity can be used for reactive power generation, $q_{\text{avail}}$, which is input into the VV controller to determine the setpoint $u^{q}_{t}$.  We note that this dynamic model of smart inverter behavior is consistent with IEEE 1547 and supporting documents \cite{inverter2016, IEEE_1547}.

The intelligent agent interacts directly with the VV/VW controllers through an action taken at time $t$ ($a_{t}$).  The subsequent reward $r_{t}$ due to the application of $a_{t}$ is then input into the agent, along with a set of observations $o_{t}$ from the grid as well as past actions $a_{t-1}$.

%%%%%%%%%%%%%%%%%%%%%%%%%%%%%%%%%%%%%%%%%%%%
\subsection{Observations of Voltage Oscillations and Imbalances}

During a cyber-attack, we assume that an adversary has the capability to alter the parameter vector $\bm{\eta}$ in a portion of DER thereby adjusting the shape of their VV/VW curves to generate large voltage oscillations \cite{roberts2020deep} or large voltage imbalances \cite{roberts2021deep_unbalance}.  We then assume that the remaining non-compromised DER in the system are equipped with an intelligent agent depicted in Fig. \ref{fig:inverter_block_diagram} which will re-adjust the parameter vector $\bm{\eta}$ to mitigate the cyber-attack in real time.

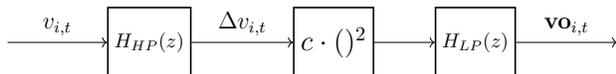
\begin{figure}[ht]
\centering
\resizebox {21pc} {!} { \tikzstyle{block} = [draw, thick, rectangle, 
    minimum height=3.5em, minimum width=4em]
\tikzstyle{triang} = [draw, thick, isosceles triangle, minimum height = 3em,
        isosceles triangle apex angle=60]
\tikzstyle{input} = [coordinate]
\tikzstyle{output} = [coordinate]
\tikzstyle{pinstyle} = [pin edge={to-,thin,black}]

\begin{tikzpicture}[auto, node distance=2.5cm]

    \node [input, name=input] {};
    \node [block, right of=input] (hpf) %{\LARGE{$\frac{s}{s+h}$}};
    {$H_{HP}(z)$};
    \node [block, right of=hpf,
            node distance=3.25cm] (square_sig) {\Large{$c\cdot ( \text{} )^{2}$}};
    % \node [triang, right of=hpf,
    %         node distance=2.25cm] (square_sig) {\Large{$k\bullet ( \text{} )^{2}$}};
    \node [block, right of=square_sig,
            node distance=2.5cm] (lpf) %{\LARGE{$\frac{l}{s+l}$}};
            {$H_{LP}(z)$};

    \draw [draw,->] (input) -- node {\large{$v_{i,t}$}} (hpf);
    \draw [->] (hpf) -- node[name=u1] {\large{$\Delta v_{i,t}$}} (square_sig);
    \draw [->] (square_sig) -- node[name=u2] {} (lpf);
    \node [output, right of=lpf] (output) {};
    \draw [->] (lpf) -- node [name=y] {\large{$\mathbf{vo}_{i,t}$}}(output);

\end{tikzpicture} }
\caption{Block diagram of illustrating the filtering process used to compute a measurement of the intensity of voltage oscillations.}
\label{fig:observer}
\end{figure}

Critical observations needed by the agent for cyber-attack mitigation are measurements of the intensity of the voltage oscillations and imbalances computed from grid telemetry.  We refer the reader to our previous works \cite{roberts2020deep, roberts2021deep_unbalance} for detailed discussions of how these observations are computed in real time.  The procedure is briefly summarized here:

\textbf{Voltage Oscillations (VO):} We utilize an intuitive filtering process to extract the ``energy" associated with observed voltage oscillations.  The filter consists of the series connection of a high-pass filter $H_{HP}$, a signal square element (with positive gain $c$), and a low-pass filter $H_{LP}$, shown in Fig. \ref{fig:observer}.  The output of the filter $\mathbf{vo}_{i,t}$ is a non-negative value which becomes larger as the amplitude of the oscillations in node $i$ voltage ($v_{i,t}$) increase.  For proper operation, the high and low-pass filter critical frequencies should be chosen as to not attenuate oscillations resulting from cyber-attacked inverters.

\textbf{Voltage Imbalances (VI):} The metric used to compute voltage imbalance at node $i$ at time $t$ is given by:
\begin{align}\label{eq:vi}
    \text{vi}_{i,t} = \dfrac{\max(|\bar{v}_{i, t} - v^{a}_{i, t}|, |\bar{v}_{i, t} - v^{b}_{i, t}|, |\bar{v}_{i, t} - v^{c}_{i, t}|)}{\bar{v}_{i, t}}
\end{align}
where $\bar{v}_{i}$ denotes the mean measured voltage magnitude at bus $i$, and $v^{a}_{i, t}$, $v^{b}_{i, t}$, $v^{c}_{i, t}$ are the measured voltage magnitudes on phase $a$, $b$, and $c$ respectively.

%%%%%%%%%%%%%%%%%%%%%%%%%%%%%%%%%%%%%%%%%%%%
\subsection{Adam}

Adam is a first-order gradient-based stochastic optimization method that which utilizes adaptive estimates of lower-order moments \cite{kingma2017adam}.  Similar to AdaGrad \cite{duchi2011adaptive} and RMSProp \cite{hintonRMS}, the method features an adaptive learning rate and has been shown to work well with sparse gradients and in non-stationary settings.  We present the Adam algorithm for completeness in Algorithm 1.

\SetKwInput{KwHyperparameters}{Hyperparameters}
\SetKwInput{KwInitialize}{Initialize}
\SetKwFunction{FADAM}{ADAM}
\SetKwProg{Fn}{Function}{:}{}
\begin{algorithm2e*}
\caption{Adam Optimization Algorithm 1-step forward}
\SetAlgoLined
\LinesNumbered
\DontPrintSemicolon
\KwHyperparameters{Gradient $g_t$, stepsize $\alpha$, exponential decay rate ${\beta}_0, {\beta}_1$ for moment estimates, tolerance parameter $\lambda_{ADAM} > 0$ for numerical stability. $m_0, v_0 \leftarrow [0, 0, 0]$}
\Fn{\FADAM{$\theta_{j}, g_j, \alpha, {\beta}_0, {\beta}_1$}}
{
$m_j \leftarrow \beta_1 \cdot m_{j-1} + (1 - \beta_1) \cdot g_j$ \;
$v_j \leftarrow \beta_2 \cdot v_{j-1} + (1 - \beta_1) \cdot g_j ^ 2$ \;
$\hat{m_j} \leftarrow m_j/(1-\beta^j_1)$ \;
$\hat{v_j} \leftarrow v_j/(1-\beta^j_2)$ \;
$\theta_{j+1} \leftarrow \theta_{j} - \alpha \cdot \hat{m_j}/(\sqrt{\hat{v_j}} + \lambda_{ADAM})$ \;
\KwRet{$\theta_t$}
}
\end{algorithm2e*}

%%%%%%%%%%%%%%%%%%%%%%%%%%%%%%%%%%%%%%%%%%%%
\subsection{Random Search and Augmented Random Search}

The goal of reinforcement learning is to determine a policy that governs a dynamic system to maximize a reward associated with a specific task.  This problem can be formulated as \cite{mania2018simple}:
\begin{equation} \underset{{\theta \in \mathbb{R}^n}}{\max}\text{ }\mathbb{E}_{\xi}[r(\pi_\theta, \xi)],
\label{eq:MDP}
\end{equation}
\noindent where $\pi_\theta$ is the policy parameterized by $\theta \in \mathbb{R}^n$, $\xi$ represents the randomness of the environment, $r(\pi_\theta, \xi)$ is the cumulative reward that the policy $\pi_\theta$ generates in one trajectory (i.e., a "rollout") of the system.

In situations where the gradient of the objective function is not directly available to aid in the search for for the optimal parameter vector $\theta^{*}$, one can utilize \emph{measurements} of the reward to approximate the gradient of the reward under the present policy $g(\theta)$.  A \emph{finite-difference} approximation of $g(\theta)$ is given by \cite{spall2003stochastic}:
\begin{equation}
    g(\theta) = \frac{r(\pi_{\theta + c\delta}, \xi_{1}) - r(\pi_{\theta - c\delta}, \xi_{2})}{2c}, \label{eq:g}
\end{equation}
\noindent where $c$ is a positive scalar and $\delta$ is a zero-mean and Gaussian vector. Random Search (RS) \cite{salimans2017evolution, spall2003stochastic, mania2018simple} incorporates single or mini-batches of gradient approximations of \eqref{eq:g} into simple gradient ascent to search for $\theta^{*}$.

%Random search (RS) is a derivative-free optimization method. Let $\bm{s} \in \mathbb{R}^{d-1}$ is a uniformly random direction on the sphere in the parameters space and $\delta$ is a small perturbation. The gradient estimator is defined as:
%\[g = \frac{d}{2\delta}\mathbb{E}_{\xi, \bm{s} \in \mathbb{R}^{d-1}}[y(\theta, \bm{s})]\]
%where
%\[y(\theta, \bm{s}) = (R(\theta +\delta \bm{s}, \xi) - R(\theta - \delta \bm{s}, \xi))\bm{s}\]

Mania et. al. \cite{mania2018simple} suggest several improvements to RS, yielding the Augmented Random Search (ARS) algorithm.  The modifications are:
\begin{itemize}
  \item Normalization of the states
  \item Scaling the gradient by the standard deviation of return
  \item Using top performing directions in mini-batch updates 
\end{itemize}

\section{Methodology}
\label{sec:methodology}

%%%%%%%%%%%%%%%%%%%%%%%%%%%%%%%%%%%%%%%%%%%
\subsection{Adam-based Augmented Random Search}

We propose an extension of ARS based on the replacement of vanilla gradient ascent with the Adam optimization algorithm.  The method is specified in Algorithm 2.

\SetKwInput{KwHyperparameters}{Hyperparameters}
\SetKwInput{KwInitialize}{Initialize}
\begin{algorithm2e*}
\SetAlgoLined
\LinesNumbered
\DontPrintSemicolon
\KwHyperparameters{number of directions sampled per iteration $N$, exploration noise $\nu$, number of top-performing directions $b$ $(b \le N)$}
\KwInitialize{ 
$\mu_0 = \mathbf{0} \in \mathbb{R}^n, \Sigma_0=\mathbf{I}_n \in \mathbb{R}^{n \times n}, j = 0$ \
\begin{equation*}
 \pi_{\theta} =
    \begin{cases}
      \mathbf{0} \in \mathbb{R}^{p \times n}  \text{ if using linear policy}\\
      \operatorname{NN}(\mathbf{0}) \text{ if using non-linear policy}\\
    \end{cases} 
\end{equation*}
}

\While{ending condition not satisfied}{
\text{Sample } $\delta_1, \delta_2, ..., \delta_N$ of appropriate dimension with i.i.d. standard normal entries. \;

Collect $2N$ rollouts of horizon $H$ and their corresponding rewards using the $2N$ policies. 
\begin{equation*}
    \begin{aligned}
        \pi_{\theta_j, k, +}(\tilde{x}) = \pi_{\theta_j + \nu\delta_k}&(\tilde{x}) \quad \text{and} \quad\pi_{\theta_j, k, -}(\tilde{x}) = \pi_{\theta_j - \nu\delta_k}(\tilde{x})  \\
        \text{where }\tilde{x} &= \operatorname{diag}(\Sigma_j)^{\frac{-1}{2}}(x-\mu_j)  
    \end{aligned}
\end{equation*}
$\operatorname{for} k \in \{1,2,3,...N\}$. \;
Sort the direction $\delta_k$ by $\text{max}\{r(\pi_{\theta_j,k,+}), r(\pi_{\theta_j,k,-})\}$, denote by $\delta_{(k)}$ the $k$-th largest direction, and by $\pi_{\theta_j,(k),+}$ and $\pi_{\theta_j,(k),-}$ the corresponding policies. \;
Policy update step: \
\begin{equation*}
    \begin{aligned}
        g_j &= \frac{\alpha}{b\sigma_R}\sum^{b}_{k=1}[r(\pi_{\theta_j,(k),+}) - r(\pi_{\theta_j,(k),-})]\delta_{(k)} \\
        \theta_{j+1} &= \operatorname{ADAM}(\theta_j, g_j, \alpha, \beta_0, \beta_1) 
    \end{aligned}
\end{equation*}
where $\sigma_R$ is the standard deviation of the $2b$ directions used to update step. \; 
Set $\mu_{j+1}, \Sigma_{j+1}$ to be mean and the covariance of the $2NH(j+1)$ states encountered from the start of training. \;
$j \leftarrow j + 1$
}
\caption{ADAM-based Augmented Random Search}
\end{algorithm2e*}

%%%%%%%%%%%%%%%%%%%%%%%%%%%%%%%%%%%%%%%%%%%
\subsection{Training Framework}

We seek to train a policy to mitigate large voltage oscillations (VO) and voltage imbalances (VI) stemming from DER smart inverter VV/VW control functions with maliciously chosen setpoints.  We represent a single distribution feeder as a graph $G = (\mathcal{N}, \mathcal{L})$, where $\mathcal{N}$ and $\mathcal{L}$ denote the set of nodes and lines, respectively.  For simplicity of presentation, it is assumed that a DER equipped with VV/VW functionality is located at every node in $\mathcal{G}$.  We now partition the set of inverters into two groups  $\mathcal{H}$ and $\mathcal{U}$ representing "compromised" and "non-compromised" sets of DER, where $\mathcal{H} \bigcup \mathcal{U} = \mathcal{N}$.  We also make the assumption that the set $\mathcal{U}$ is nonempty and contains sufficient amounts of DER to mitigate the effect of the cyber-attack.  We adopt the following framework for training optimal policies:
%The goal of the controller is to mitigate system instabilities, Voltage Unbalance (VU) and Voltage Oscillation (VO), caused by DER smart inverter VV/VW maliciously chosen setpoints.   Let presents the distribution feeder as a graph $G = (\mathcal{N}, \mathcal{L})$, where $\mathcal{N}$ is the set of nodes and $\mathcal{L}$ is the set of lines. For simplicity, we assume that a VV/VW smart inverter is installed a t every node in the feeder, so the total number of inverters in the feeder is $\mathcal{N}$. We suppose that $\mathcal{N}$ is partitioned into to two sets, $\mathcal{H}$ and $\mathcal{U}$, which represent the "compromised" and "uncompromised" sets, so $\mathcal{H} \bigcup \mathcal{U} = \mathcal{N}$. We also assume that the controllable resources, $\mathcal{U} \neq \emptyset$, are adequate to mitigate the effects of the cyber-physical attack. 
%

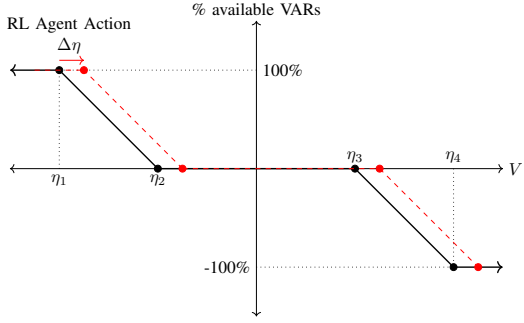
\begin{figure}[H]
    \centering
    \resizebox {17pc} {!} { \begin{tikzpicture}

% horizontal axis
\coordinate (xmax) at (5,0);
\coordinate (xmin) at (-5,0);
\draw[<->] (xmin) -- (xmax) node[anchor=west] {$V$};
    
% vertical axis
\coordinate (ymax) at (0,3);
\coordinate (ymin) at (0,-3);
\draw[<->] (ymin) -- (ymax) node[anchor=south] {\% available VARs};
    
% labels
\filldraw[black]	
		(-2,0) circle (2pt) node[anchor=north] {$\eta_{2}$}
		(2,0) circle (2pt) node[anchor=south] {$\eta_{3}$}
        
        (-4,2) circle (2pt)
        (4,-2) circle (2pt);
        
\draw	
		(0,2) node[anchor=west] {100\%}
		(-4,0) node[anchor=north] {$\eta_{1}$}
		(0,-2) node[anchor=east] {-100\%}
		(4,0) node[anchor=south] {$\eta_{4}$};

\draw[dotted] (4,0) -- (4,-2);
\draw[dotted] (-4,0) -- (-4,2);

\draw[dotted] (0,-2) -- (4,-2);
\draw[dotted] (0,2) -- (-4,2);

% curve
\draw[thick][<->] (-5,2) -- (-4,2) -- (-2,0) -- (2,0) -- (4,-2) -- (5,-2);

% labels
\filldraw[red]	
		(-1.5,0) circle (2pt)
		(2.5,0) circle (2pt)
        (-3.5,2.0) circle (2pt)
        (4.5,-2.0) circle (2pt);

\draw[red, dashed] (-1.5, 0) -- (2.5,0);
\draw[red, dashed] (-1.5, 0) -- (-3.5,2);
\draw[red, dashed] (2.5, 0) -- (4.5,-2);
\draw[red, dashed] (-4.5, 2) -- (-3.5,2);

\draw(-3.8,3.2) node[anchor=north, align=center] {RL Agent Action\\$\Delta \eta$};
\draw[red, ->] (-4,2.2) -- (-3.5,2.2);
%\draw[red, dashed] (-1.5, 0) -- (2.5,0);
%\draw[dotted] (-4,0) -- (-4,2);

%\draw[dotted] (0,-2) -- (4,-2);
%\draw[dotted] (0,2) -- (-4,2);
% \draw	
% 		(0,2) node[anchor=west] {100\%}
% 		(-4,0) node[anchor=north] {$v_{1}$}
% 		(0,-2) node[anchor=east] {-100\%}
% 		(4,0) node[anchor=south] {$v_{4}$};

% \draw[dotted] (4,0) -- (4,-2);
% \draw[dotted] (-4,0) -- (-4,2);

% \draw[dotted] (0,-2) -- (4,-2);
% \draw[dotted] (0,2) -- (-4,2);

% % curve
% \draw[thick][<->] (-5,2) -- (-4,2) -- (-2,0) -- (2,0) -- (4,-2) -- (5,-2);

\end{tikzpicture}}
    \caption{Action example.}
    \label{fig:example_action}
\end{figure}

\noindent{\bf Training}: 
\begin{enumerate}
    \item We define a single agent whose input observation vector is from a single node in the network at time $t$ with worst case VU and VO. The agent has a multi-head output action, $a^{i}_{t}\forall i \in \{a,b,c\}$, which is a deviation/offset, $\Delta \eta$, from default VV/VW control curves in Figs. \ref{fig:vvc} - \ref{fig:vwc} that is applied across all single-phase inverters. An example of an action is shown in Fig. \ref{fig:example_action}.
    The action space is a continuous offset between -0.1 pu to 0.1 pu around the default inverter VV/VW curve.
    \item In order to preserve the Markov property, new parameterizations of VV/VW functions occurs on a slower timescale than the filter dynamics in Fig. \ref{fig:inverter_block_diagram}.
\end{enumerate}

{\bf Observation}: The observation vectors $o_{i,t}$, where $i\in \cal U$, at each controller consist of:
\begin{enumerate}
    \item $\text{vi}_{t}$: defined in \eqref{eq:vi}.
    \item $\text{vo}_{t}$: defined in Fig. \ref{fig:observer}.
    \item $q^{\text{avail, nom}}_{t}$: the available reactive power capacity without active power curtailment.
    \item $a^\text{a}_{t-1}$, $a^\text{b}_{t-1}$, $a^\text{c}_{t-1}$: the previous action taken by the agent across each phase.
    \item $v^\text{a}_{t}$, $v^\text{b}_{t}$, $v^\text{c}_{t}$: voltage phase measurements at time $t$
\end{enumerate} 

During the training of the agent, $\text{vo}_{t}$ and $\text{vu}_{t}$ are the \emph{worst-case} oscillations and imbalances in the feeder, and $q^{\text{avail, nom}}_{t}$ is the average across all non-compromised inverters. When the agent is deployed to manage individual DER, $\text{vo}_{t}$ and $\text{vu}_{t}$ will be collected from the nearest three-phase node. 

{\bf Reward}: At a time $t$, the reward function, $r_{t}(a_{t},o_t)$ for a single agent in training is:
\begin{equation}\label{eq:rt}
\begin{split}
r_{t} = &-\Bigg(\sigma_{u} \vert\vert \bm{\text{vi}}_{t} \vert\vert_{\infty} + \sigma_{u} \vert\vert \bm{\text{vo}}_{t} \vert\vert_{\infty} + \sum_{i\in\{a,b,c\}}\sigma_{a} \boldsymbol{1}_{a^i_t\neq a^i_{t-1}} + \\ &\sum_{i\in\{a,b,c\}} \sigma_{0} \lVert a^i_t \rVert_{2} + \frac{1}{|\mathcal{U}|}\sum_{j=1}^{|\mathcal{U}|}\sigma_{p} \left(1 - \frac{p_{j,t}}{p_{j,t}^{\text{max}}}\right)^2 \Bigg).
\end{split}
\end{equation}
The first term seeks to minimize the maximum imbalance, $\vert\vert \bm{\text{vi}}_{t} \vert\vert_{\infty}$ and the second term seeks to minimize the maximum oscillation, $\vert\vert \bm{\text{vu}}_{t} \vert\vert_{\infty}$ over all nodes in the network. For details on the remaining components of the reward, we refer the reader to our previous work \cite{roberts2021deep_unbalance}. 

\section{Results}
\label{sec:results}

\subsection{Experimental Setup}
Experiments were conducted on the IEEE-37 node test feeder where DER with VV/VW capability were placed at all load buses with peak active power generation of 100\% of nominal load.  Additionally, the inverter capacity associated with all DER was oversized by 10\% to allow for some reactive power compensation without active power curtailment at all simulation timesteps.  Agent training occurs using 700 second rollouts with 1 second timestep simulations using OpenDSS.  Load, solar generation, percentage of DER resource which are compromised, and the phase of the voltage regulator are all randomized for each rollout.  At $t=200$ seconds in the simulation, a simulated attack is launched where the attacker controls between 10\% and 40\% of inverter capacity in the system.

\subsection{Training performance}
\begin{figure}
\centering
\def\ars{data/reward_convergence_ars.csv}
\def\arsadam{data/reward_convergence_ars_adam.csv}
\def\ppo{data/reward_convergence_ppo.csv}
\def\ymin{-150}
\newcommand{\errorband}[5][]{
\pgfplotstableread[col sep=comma]{#2}\datatable
  \addplot [name path=pluserror,draw=none,no markers,forget plot]
    table [x={#3},y expr=\thisrow{#4}+\thisrow{#5}] {\datatable};

  \addplot [name path=minuserror,draw=none,no markers,forget plot]
    table [x={#3},y expr=\thisrow{#4}-\thisrow{#5}] {\datatable};

  \addplot [forget plot, fill=gray!40, #1]
    fill between[on layer={},of=pluserror and minuserror];
}

\begin{tikzpicture}
\definecolor{tableau_blue}{RGB}{31,119,180}
\definecolor{tableau_orange}{RGB}{255,127,14}
\definecolor{tableau_green}{RGB}{44,160,44}

\begin{groupplot}[group style={
        group name=my plots,
        group size=1 by 2,
        xlabels at=edge bottom,
        vertical sep=16pt,
    },
    xlabel={Epochs},
    legend style={nodes={scale=0.6, transform shape}},
    width=\linewidth,
    height=3.8 cm,
    enlarge x limits=false,
    tick pos=left,
    ymajorgrids=true,
    xmajorgrids=true,
    yminorgrids=false,
    xticklabel style={font=\scriptsize},
    yticklabel style={font=\scriptsize,rotate=90,anchor=base,yshift=0.2cm,scaled y ticks=false, /pgf/number format/fixed},
    minor y tick num=0,
]

\nextgroupplot[ylabel=reward,legend pos=south east, ymax=-80, ymin=-140, ytick={-140, -110 ,-90}, axis x line*=top]
\errorband[tableau_blue, opacity=0.4]{\arsadam}{}{mean}{std}
\addplot [smooth, thick, tableau_blue, no markers] table [x=, y=mean, col sep=comma] {\arsadam};
\errorband[tableau_orange, opacity=0.4]{\ars}{}{mean}{std}
\addplot [smooth, thick, tableau_orange, no markers] table [x=, y=mean, col sep=comma] {\ars};
\addlegendentry{Adam ARS}
\addlegendentry{ARS}

\nextgroupplot[ylabel=reward,legend pos=south east, ymax=-80, ymin=-500, ytick={-500, -300, -90}]
\errorband[tableau_blue, opacity=0.4]{\arsadam}{}{mean}{std}
\addplot [smooth, thick, tableau_blue, no markers] table [x=, y=mean, col sep=comma] {\arsadam};
\errorband[tableau_orange, opacity=0.4]{\ars}{}{mean}{std}
\addplot [smooth, thick, tableau_orange, no markers] table [x=, y=mean, col sep=comma] {\ars};
\errorband[tableau_green, opacity=0.4]{\ppo}{}{mean}{std}
\addplot [smooth, thick, tableau_green, no markers] table [x=, y=mean, col sep=comma] {\ppo};
\addlegendentry{Adam ARS}
\addlegendentry{ARS}
\addlegendentry{PPO}

\end{groupplot}
\end{tikzpicture}
\caption{Average training reward. The shaded area represents the standard deviation over 10 runs.}
\label{fig:reward}
\end{figure}
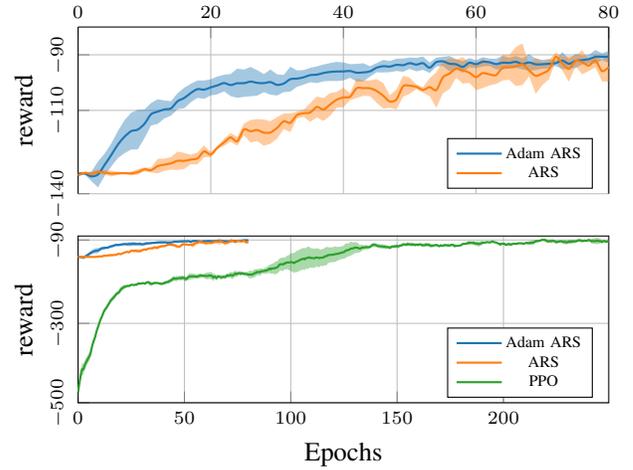

Fig. \ref{fig:reward} shows the training performance of Adam-ARS, ARS, and PPO when these agents are learning to mitigate a voltage imbalance attack. All three algorithms utilize the same neural network architecture. As is shown in the figures, ARS and ADAM-based ARS take approximately 80 epochs to converge to the optimal policy while PPO takes more than 3 times the number of epochs to converge. Interestingly, due to the nature of training a deep neural network with backpropagation \cite{pmlr-v9-glorot10a}, the weights of neural network in PPO requires fine-tuned initialization. This initialization results in extremely poor performance of PPO at the start of training, taking over 100 epochs to achieve rewards similar to both ARS variants. In contrast, since ARS/Adam-ARS randomly perturbs the parameters of the neural network, neural networks for these algorithms can be initialized with weights equal to 0.

The top subplot of Figure \ref{fig:reward} shows Adam-ARS learns the optimal policy faster than ARS and shows much less variance in the associated rewards. 

\subsection{Oscillation Attacks}
\begin{figure}[ht!]
\centering
\def\evalhist{data/ars_h30_t100_oscillation.csv}
\def\hackstart{200}
\def\hackend{450}
\begin{tikzpicture}

% how to generate csv: see understand_hyperparameters.py

% from https://tableaufriction.blogspot.com/2012/11/finally-you-can-use-tableau-data-colors.html  
\definecolor{tableau_blue}{RGB}{31,119,180}
\definecolor{tableau_orange}{RGB}{255,127,14}
\definecolor{tableau_red}{RGB}{214,39,40}
\definecolor{tableau_grey}{RGB}{127,127,127}
\definecolor{tableau_green}{RGB}{44,160,44}
\definecolor{tableau_purple}{RGB}{148,103,189}
\definecolor{tableau_cyan}{RGB}{23,190,207}

% vertical lines right and left of hack (not used)
\newcommand{\hackstartline}{
\draw [tableau_red, very thick, dashed] 
(axis cs:\hackstart,\pgfkeysvalueof{/pgfplots/ymin}) -- 
(axis cs:\hackstart,\pgfkeysvalueof{/pgfplots/ymax});}

\newcommand{\hackendline}{
\draw [tableau_red, very thick, dashed] 
(axis cs:\hackend,\pgfkeysvalueof{/pgfplots/ymin}) -- 
(axis cs:\hackend,\pgfkeysvalueof{/pgfplots/ymax});}

\newcommand{\hacklines}{\hackstartline\hackendline}

% shaded area over hack
\newcommand{\hackshade}{
\draw [fill=tableau_red, opacity=.06] 
(axis cs:\hackstart,\pgfkeysvalueof{/pgfplots/ymin}) -- 
(axis cs:\hackend,\pgfkeysvalueof{/pgfplots/ymin}) -- 
(axis cs:\hackend,\pgfkeysvalueof{/pgfplots/ymax}) -- 
(axis cs:\hackstart,\pgfkeysvalueof{/pgfplots/ymax});
}

\begin{groupplot}[group style={
        group name=my plots,
        group size=1 by 4,
        xlabels at=edge bottom,
        xticklabels at=edge bottom,
        vertical sep=15pt,
    },
    xlabel={Time (s)},
    legend style={nodes={scale=0.6, transform shape}},
    width=\linewidth,
    height=3.3cm,
    enlarge x limits=false,
    tick pos=left,
    xticklabel style={font=\scriptsize},
    yticklabel style={font=\scriptsize,rotate=90,anchor=base,yshift=0.2cm,scaled y ticks=false, /pgf/number format/fixed},
    minor y tick num=0,
    title style={yshift=-8pt}
]
%\nextgroupplot[ylabel=voltage,legend pos=north east, ymax=1.06, ymin=0.96,try min ticks=3]\hackshade
\nextgroupplot[ylabel=voltage (p.u.),legend pos=south west, ymax=1.02, ymin=0.92, ytick={0.93, 0.97, 1.02},legend columns=1]\hackshade
\addplot [smooth, tableau_blue, no markers] table [x=, y=va, col sep=comma] {\evalhist};
\addplot [smooth, tableau_green, no markers] table [x=, y=vb, col sep=comma] {\evalhist};
\addplot [smooth, tableau_cyan, no markers] table [x=, y=vc, col sep=comma] {\evalhist};
\addlegendentry{Phase A}
\addlegendentry{Phase B}
\addlegendentry{Phase C}
\coordinate (hackstart) at (axis cs:\hackstart,\pgfkeysvalueof{/pgfplots/ymax});
\coordinate (hackend) at (axis cs:\hackend,\pgfkeysvalueof{/pgfplots/ymax});

%\hacklines
%\addlegendentry{voltage}

%\nextgroupplot[ylabel=VU,legend pos=north east, ymin=-0.005, ymax=0.06, ytick={0, 0.02, 0.04}]\hackshade
\nextgroupplot[ylabel=VO, legend pos=north east, ymin=0, ymax=0.5, ytick={0, 0.2, 0.4}, title=Largest Network Voltage Oscillation]\hackshade
\addplot [tableau_red, no markers] table [x=,  y expr=\thisrow{y_worst}, col sep=comma] {\evalhist};
%\hacklines
%\addlegendentry{oscillation observer}

\nextgroupplot[ylabel=$\Delta \eta$, legend pos=north east, legend columns=4,try min ticks=1,ymin=-0.11,ymax=0.11, title=Agent action]\hackshade
\addplot [tableau_blue, no markers] table [x=, y=ta_0, col sep=comma] {\evalhist};

\addplot [tableau_green, no markers] table [x=, y=tb_0, col sep=comma] {\evalhist};

\addplot [tableau_cyan, no markers] table [x=, y=tc_0, col sep=comma] {\evalhist};

%\hacklines
\addlegendentry{Offset phase A}
\addlegendentry{Offset phase B}
\addlegendentry{Offset phase C}

% \nextgroupplot[ylabel=tap position, legend pos=north east, legend columns=4,try min ticks=1, ymin=-4, ymax=17, title=Voltage Regulator]\hackshade
% \addplot [tableau_purple, no markers] table [x=, y=reg, col sep=comma] {\evalhist};

%\addplot [tableau_red, no markers] table [x=, y=ht_all, col sep=comma] {\evalhist};
%\hacklines

% \nextgroupplot[ylabel=reward, legend pos=south east, legend columns=1, ymax=5, ymin=-50]\hackshade
% \addplot [tableau_purple, no markers] table [x=, y=component_y, col sep=comma] {\evalhist};
% \addplot [tableau_green, no markers] table [x=, y=component_oa, col sep=comma] {\evalhist};
% \addplot [tableau_orange, no markers] table [x=, y=component_init, col sep=comma] {\evalhist};
% \addplot [tableau_red, no markers] table [x=, y=component_pset_pmax, col sep=comma] {\evalhist};
% \addplot [tableau_blue, no markers] table [x=, y=total_reward, col sep=comma] {\evalhist};

% %\hacklines
% \addlegendentry{Oscillation}
% \addlegendentry{Last action}
% \addlegendentry{Initial action}
% \addlegendentry{P curtailment}
% \addlegendentry{Total reward}

% \nextgroupplot[ylabel=kVar, legend pos=south east, legend columns=4,ytick={-1000, 0, 1000}, ymin=-2000,ymax=1500, title= Total Reactive Power Injection]\hackshade
\nextgroupplot[ylabel=MVar, legend pos=south east, legend columns=4, ytick={0, 1}, ymin=-1,ymax=1.5, title= Total DER Reactive Power Injection]\hackshade
\addplot [tableau_blue, no markers] table [x=, y expr=-\thisrow{qouta_total}/1000, col sep=comma] {\evalhist};
\addplot [tableau_red, no markers] table [x=, y expr=-\thisrow{qoutb_total}/1000, col sep=comma] {\evalhist};
\addplot [tableau_green, no markers] table [x=, y expr=-\thisrow{qoutc_total}/1000, col sep=comma] {\evalhist};

%\addplot [name path=upper,draw=none] table[x=,y expr=-\thisrow{qouta_mean}+\thisrow{qouta_std}] {\evalhist};
%\addplot [name path=lower,draw=none] table[x=,y expr=-\thisrow{qouta_mean}-\thisrow{qouta_std}] {\evalhist};
%\addplot [fill=blue!10] fill between[of=upper and lower];
%\errorband[tableau_blue, opacity=0.5]{\evalhist}{}{qouta_mean}{qouta_std};
%\errorband[tableau_blue, opacity=0.4]{\evalhist}{}{mean}{std}

\addlegendentry{Phase A}
\addlegendentry{Phase B}
\addlegendentry{Phase C}

\end{groupplot}

\draw[latex-latex, thick, tableau_red, yshift=0.5cm] (hackstart) to node [auto] {attack} (hackend);

\end{tikzpicture}
\caption{30\% DER oscillation attack with ADAM-based ARS}
\label{fig:ars_vo}
\end{figure}
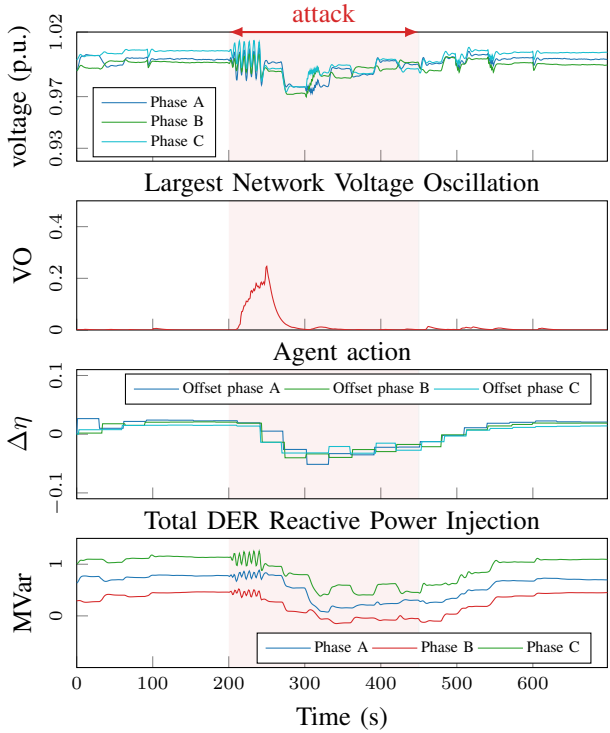

An attack on 30\% of DER VV/VW controllers that creates voltage oscillations is depicted in Fig. \ref{fig:baseline_vo}.  This baseline case demonstrates the effect of the attack in the system without utilizing the policy trained by Adam-ARS to control the remaining DER smart inverters in the system.  Fig. \ref{fig:ars_vo} shows the effect of a \emph{linear} policy trained by Adam-ARS. Clearly, the trained policy is effective in adjusting DER smart inverter VV/VW controllers to minimize voltage oscillations in the network shortly after they first manifest.  Hyperparameters for this experiment are shown in Table \ref{table:hyperparams1}.

\begin{figure}[ht!]
\centering
\def\evalhist{data/baseline_h30_t100_oscillation.csv}
\def\hackstart{200}
\def\hackend{450}
\begin{tikzpicture}

% how to generate csv: see understand_hyperparameters.py

% from https://tableaufriction.blogspot.com/2012/11/finally-you-can-use-tableau-data-colors.html  
\definecolor{tableau_blue}{RGB}{31,119,180}
\definecolor{tableau_orange}{RGB}{255,127,14}
\definecolor{tableau_red}{RGB}{214,39,40}
\definecolor{tableau_grey}{RGB}{127,127,127}
\definecolor{tableau_green}{RGB}{44,160,44}
\definecolor{tableau_purple}{RGB}{148,103,189}
\definecolor{tableau_cyan}{RGB}{23,190,207}

% vertical lines right and left of hack (not used)
\newcommand{\hackstartline}{
\draw [tableau_red, very thick, dashed] 
(axis cs:\hackstart,\pgfkeysvalueof{/pgfplots/ymin}) -- 
(axis cs:\hackstart,\pgfkeysvalueof{/pgfplots/ymax});}

\newcommand{\hackendline}{
\draw [tableau_red, very thick, dashed] 
(axis cs:\hackend,\pgfkeysvalueof{/pgfplots/ymin}) -- 
(axis cs:\hackend,\pgfkeysvalueof{/pgfplots/ymax});}

\newcommand{\hacklines}{\hackstartline\hackendline}

% shaded area over hack
\newcommand{\hackshade}{
\draw [fill=tableau_red, opacity=.06] 
(axis cs:\hackstart,\pgfkeysvalueof{/pgfplots/ymin}) -- 
(axis cs:\hackend,\pgfkeysvalueof{/pgfplots/ymin}) -- 
(axis cs:\hackend,\pgfkeysvalueof{/pgfplots/ymax}) -- 
(axis cs:\hackstart,\pgfkeysvalueof{/pgfplots/ymax});
}

\begin{groupplot}[group style={
        group name=my plots,
        group size=1 by 4,
        xlabels at=edge bottom,
        xticklabels at=edge bottom,
        vertical sep=15pt,
    },
    xlabel={Time (s)},
    legend style={nodes={scale=0.6, transform shape}},
    width=\linewidth,
    height=3.3cm,
    enlarge x limits=false,
    tick pos=left,
    xticklabel style={font=\scriptsize},
    yticklabel style={font=\scriptsize,rotate=90,anchor=base,yshift=0.2cm,scaled y ticks=false, /pgf/number format/fixed},
    minor y tick num=0,
    title style={yshift=-8pt}
]
%\nextgroupplot[ylabel=voltage,legend pos=north east, ymax=1.06, ymin=0.96,try min ticks=3]\hackshade
\nextgroupplot[ylabel=voltage (p.u.),legend pos=south west, ymax=1.02, ymin=0.92, ytick={0.93, 0.97, 1.02},legend columns=1]\hackshade
\addplot [smooth, tableau_blue, no markers] table [x=, y=va, col sep=comma] {\evalhist};
\addplot [smooth, tableau_green, no markers] table [x=, y=vb, col sep=comma] {\evalhist};
\addplot [smooth, tableau_cyan, no markers] table [x=, y=vc, col sep=comma] {\evalhist};
\addlegendentry{Phase A}
\addlegendentry{Phase B}
\addlegendentry{Phase C}
\coordinate (hackstart) at (axis cs:\hackstart,\pgfkeysvalueof{/pgfplots/ymax});
\coordinate (hackend) at (axis cs:\hackend,\pgfkeysvalueof{/pgfplots/ymax});

%\hacklines
%\addlegendentry{voltage}

%\nextgroupplot[ylabel=VU,legend pos=north east, ymin=-0.005, ymax=0.06, ytick={0, 0.02, 0.04}]\hackshade
\nextgroupplot[ylabel=VO, legend pos=north east, ymin=0, ymax=0.5, ytick={0, 0.2, 0.4}, title=Largest Network Voltage Oscillation]\hackshade
\addplot [tableau_red, no markers] table [x=,  y expr=\thisrow{y_worst}, col sep=comma] {\evalhist};
%\hacklines
%\addlegendentry{oscillation observer}

\nextgroupplot[ylabel=$\Delta \eta$, legend pos=north east, legend columns=4,try min ticks=1,ymin=-0.11,ymax=0.11, title=Agent action]\hackshade
\addplot [tableau_blue, no markers] table [x=, y=ta_0, col sep=comma] {\evalhist};

\addplot [tableau_green, no markers] table [x=, y=tb_0, col sep=comma] {\evalhist};

\addplot [tableau_cyan, no markers] table [x=, y=tc_0, col sep=comma] {\evalhist};

%\hacklines
\addlegendentry{Offset phase A}
\addlegendentry{Offset phase B}
\addlegendentry{Offset phase C}

% \nextgroupplot[ylabel=tap position, legend pos=north east, legend columns=4,try min ticks=1, ymin=-4, ymax=17, title=Voltage Regulator]\hackshade
% \addplot [tableau_purple, no markers] table [x=, y=reg, col sep=comma] {\evalhist};

%\addplot [tableau_red, no markers] table [x=, y=ht_all, col sep=comma] {\evalhist};
%\hacklines

% \nextgroupplot[ylabel=reward, legend pos=south east, legend columns=1, ymax=5, ymin=-50]\hackshade
% \addplot [tableau_purple, no markers] table [x=, y=component_y, col sep=comma] {\evalhist};
% \addplot [tableau_green, no markers] table [x=, y=component_oa, col sep=comma] {\evalhist};
% \addplot [tableau_orange, no markers] table [x=, y=component_init, col sep=comma] {\evalhist};
% \addplot [tableau_red, no markers] table [x=, y=component_pset_pmax, col sep=comma] {\evalhist};
% \addplot [tableau_blue, no markers] table [x=, y=total_reward, col sep=comma] {\evalhist};

% %\hacklines
% \addlegendentry{Oscillation}
% \addlegendentry{Last action}
% \addlegendentry{Initial action}
% \addlegendentry{P curtailment}
% \addlegendentry{Total reward}

% \nextgroupplot[ylabel=kVar, legend pos=south east, legend columns=4,ytick={-1000, 0, 1000}, ymin=-2000,ymax=1500, title= Total Reactive Power Injection]\hackshade
\nextgroupplot[ylabel=MVar, legend pos=south east, legend columns=4, ytick={0, 1}, ymin=-1,ymax=1.5, title= Total DER Reactive Power Injection]\hackshade
\addplot [tableau_blue, no markers] table [x=, y expr=-\thisrow{qouta_total}/1000, col sep=comma] {\evalhist};
\addplot [tableau_red, no markers] table [x=, y expr=-\thisrow{qoutb_total}/1000, col sep=comma] {\evalhist};
\addplot [tableau_green, no markers] table [x=, y expr=-\thisrow{qoutc_total}/1000, col sep=comma] {\evalhist};

%\addplot [name path=upper,draw=none] table[x=,y expr=-\thisrow{qouta_mean}+\thisrow{qouta_std}] {\evalhist};
%\addplot [name path=lower,draw=none] table[x=,y expr=-\thisrow{qouta_mean}-\thisrow{qouta_std}] {\evalhist};
%\addplot [fill=blue!10] fill between[of=upper and lower];
%\errorband[tableau_blue, opacity=0.5]{\evalhist}{}{qouta_mean}{qouta_std};
%\errorband[tableau_blue, opacity=0.4]{\evalhist}{}{mean}{std}

\addlegendentry{Phase A}
\addlegendentry{Phase B}
\addlegendentry{Phase C}

\end{groupplot}

\draw[latex-latex, thick, tableau_red, yshift=0.5cm] (hackstart) to node [auto] {attack} (hackend);

\end{tikzpicture}
\caption{30\% DER oscillation attack with no defense}
\label{fig:baseline_vo}
\end{figure}
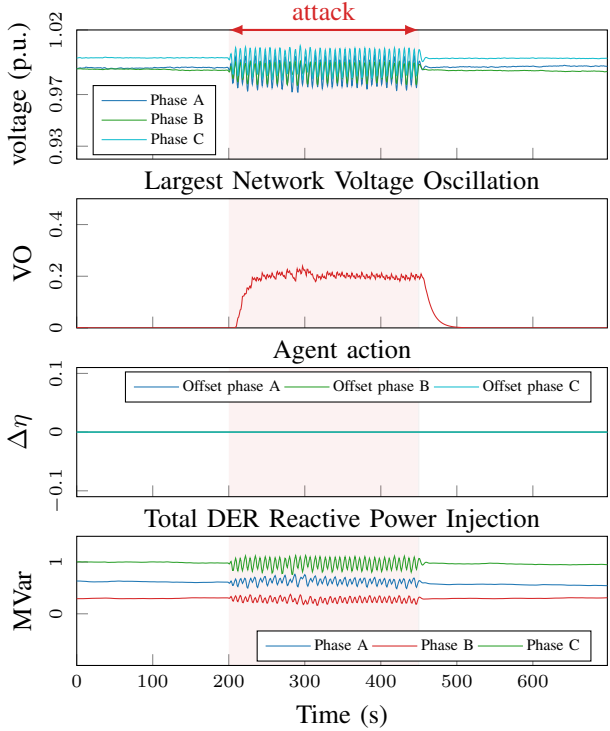

\subsection{Imbalance Attacks}
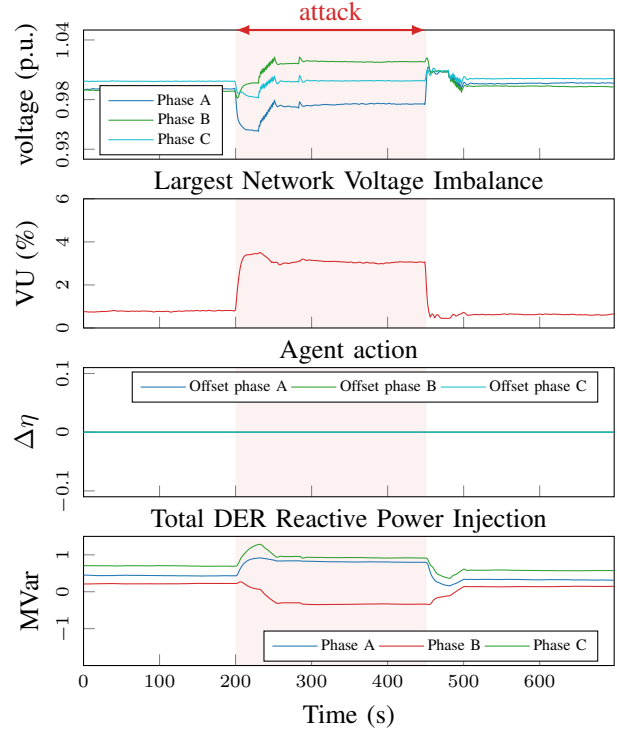
\begin{figure}[ht!]
\centering
\def\evalhist{data/baseline_h30_t100_test_distributed.csv}
\def\hackstart{200}
\def\hackend{450}
\begin{tikzpicture}

% how to generate csv: see understand_hyperparameters.py

% from https://tableaufriction.blogspot.com/2012/11/finally-you-can-use-tableau-data-colors.html  
\definecolor{tableau_blue}{RGB}{31,119,180}
\definecolor{tableau_orange}{RGB}{255,127,14}
\definecolor{tableau_red}{RGB}{214,39,40}
\definecolor{tableau_grey}{RGB}{127,127,127}
\definecolor{tableau_green}{RGB}{44,160,44}
\definecolor{tableau_purple}{RGB}{148,103,189}
\definecolor{tableau_cyan}{RGB}{23,190,207}

% vertical lines right and left of hack (not used)
\newcommand{\hackstartline}{
\draw [tableau_red, very thick, dashed] 
(axis cs:\hackstart,\pgfkeysvalueof{/pgfplots/ymin}) -- 
(axis cs:\hackstart,\pgfkeysvalueof{/pgfplots/ymax});}

\newcommand{\hackendline}{
\draw [tableau_red, very thick, dashed] 
(axis cs:\hackend,\pgfkeysvalueof{/pgfplots/ymin}) -- 
(axis cs:\hackend,\pgfkeysvalueof{/pgfplots/ymax});}

\newcommand{\hacklines}{\hackstartline\hackendline}

% shaded area over hack
\newcommand{\hackshade}{
\draw [fill=tableau_red, opacity=.06] 
(axis cs:\hackstart,\pgfkeysvalueof{/pgfplots/ymin}) -- 
(axis cs:\hackend,\pgfkeysvalueof{/pgfplots/ymin}) -- 
(axis cs:\hackend,\pgfkeysvalueof{/pgfplots/ymax}) -- 
(axis cs:\hackstart,\pgfkeysvalueof{/pgfplots/ymax});
}

\begin{groupplot}[group style={
        group name=my plots,
        group size=1 by 4,
        xlabels at=edge bottom,
        xticklabels at=edge bottom,
        vertical sep=15pt,
    },
    xlabel={Time (s)},
    legend style={nodes={scale=0.6, transform shape}},
    width=\linewidth,
    height=3.3cm,
    enlarge x limits=false,
    tick pos=left,
    xticklabel style={font=\scriptsize},
    yticklabel style={font=\scriptsize,rotate=90,anchor=base,yshift=0.2cm,scaled y ticks=false, /pgf/number format/fixed},
    minor y tick num=0,
    title style={yshift=-8pt}
]
%\nextgroupplot[ylabel=voltage,legend pos=north east, ymax=1.06, ymin=0.96,try min ticks=3]\hackshade
\nextgroupplot[ylabel=voltage (p.u.),legend pos=south west, ymax=1.05, ymin=0.92, ytick={0.93, 0.98, 1.04},legend columns=1]\hackshade
\addplot [smooth, tableau_blue, no markers] table [x=, y=va, col sep=comma] {\evalhist};
\addplot [smooth, tableau_green, no markers] table [x=, y=vb, col sep=comma] {\evalhist};
\addplot [smooth, tableau_cyan, no markers] table [x=, y=vc, col sep=comma] {\evalhist};
\addlegendentry{Phase A}
\addlegendentry{Phase B}
\addlegendentry{Phase C}
\coordinate (hackstart) at (axis cs:\hackstart,\pgfkeysvalueof{/pgfplots/ymax});
\coordinate (hackend) at (axis cs:\hackend,\pgfkeysvalueof{/pgfplots/ymax});

%\hacklines
%\addlegendentry{voltage}

%\nextgroupplot[ylabel=VU,legend pos=north east, ymin=-0.005, ymax=0.06, ytick={0, 0.02, 0.04}]\hackshade
\nextgroupplot[ylabel=VU (\%),legend pos=north east, ymin=0, ymax=6, ytick={0, 2, 4, 6}, title=Largest Network Voltage Imbalance]\hackshade
\addplot [tableau_red, no markers] table [x=,  y expr=\thisrow{u_worst}*100, col sep=comma] {\evalhist};
%\hacklines
%\addlegendentry{oscillation observer}

\nextgroupplot[ylabel=$\Delta \eta$, legend pos=north east, legend columns=4,try min ticks=1,ymin=-0.11,ymax=0.11, title=Agent action]\hackshade
\addplot [tableau_blue, no markers] table [x=, y=ta_0, col sep=comma] {\evalhist};
\addplot [tableau_green, no markers] table [x=, y=tb_0, col sep=comma] {\evalhist};
\addplot [tableau_cyan, no markers] table [x=, y=tc_0, col sep=comma] {\evalhist};
\addlegendentry{Offset phase A}
\addlegendentry{Offset phase B}
\addlegendentry{Offset phase C}

% \nextgroupplot[ylabel=tap position, legend pos=north east, legend columns=4,try min ticks=1, ymin=-4,ymax=17, title=Voltage Regulator]\hackshade
% \addplot [tableau_purple, no markers] table [x=, y=reg, col sep=comma] {\evalhist};
%\addplot [tableau_red, no markers] table [x=, y=ht_all, col sep=comma] {\evalhist};
%\hacklines

% \nextgroupplot[ylabel=reward, legend pos=south east, legend columns=1, ymax=5, ymin=-50]\hackshade
% \addplot [tableau_purple, no markers] table [x=, y=component_y, col sep=comma] {\evalhist};
% \addplot [tableau_green, no markers] table [x=, y=component_oa, col sep=comma] {\evalhist};
% \addplot [tableau_orange, no markers] table [x=, y=component_init, col sep=comma] {\evalhist};
% \addplot [tableau_red, no markers] table [x=, y=component_pset_pmax, col sep=comma] {\evalhist};
% \addplot [tableau_blue, no markers] table [x=, y=total_reward, col sep=comma] {\evalhist};

% %\hacklines
% \addlegendentry{Oscillation}
% \addlegendentry{Last action}
% \addlegendentry{Initial action}
% \addlegendentry{P curtailment}
% \addlegendentry{Total reward}

% \nextgroupplot[ylabel=kVar, legend pos=south east, legend columns=4,ytick={-1000, 0, 1000}, ymin=-2000,ymax=1500, title= Total Reactive Power Injection]\hackshade
\nextgroupplot[ylabel=MVar, legend pos=south east, legend columns=4, ytick={-1, 0, 1}, ymin=-2,ymax=1.5, title= Total DER Reactive Power Injection]\hackshade
\addplot [tableau_blue, no markers] table [x=, y expr=-\thisrow{qouta_total}/1000, col sep=comma] {\evalhist};
\addplot [tableau_red, no markers] table [x=, y expr=-\thisrow{qoutb_total}/1000, col sep=comma] {\evalhist};
\addplot [tableau_green, no markers] table [x=, y expr=-\thisrow{qoutc_total}/1000, col sep=comma] {\evalhist};

%\addplot [name path=upper,draw=none] table[x=,y expr=-\thisrow{qouta_mean}+\thisrow{qouta_std}] {\evalhist};
%\addplot [name path=lower,draw=none] table[x=,y expr=-\thisrow{qouta_mean}-\thisrow{qouta_std}] {\evalhist};
%\addplot [fill=blue!10] fill between[of=upper and lower];
%\errorband[tableau_blue, opacity=0.5]{\evalhist}{}{qouta_mean}{qouta_std};
%\errorband[tableau_blue, opacity=0.4]{\evalhist}{}{mean}{std}

\addlegendentry{Phase A}
\addlegendentry{Phase B}
\addlegendentry{Phase C}

\end{groupplot}

\draw[latex-latex, thick, tableau_red, yshift=0.5cm] (hackstart) to node [auto] {attack} (hackend);

\end{tikzpicture}
\caption{30\% DER imbalance attack with no defense}
\label{fig:baseline_vu}
\end{figure}

\begin{figure}[ht!]
\centering
\def\evalhist{data/ars_h30_t100.csv}
\def\hackstart{200}
\def\hackend{450}
\begin{tikzpicture}

% how to generate csv: see understand_hyperparameters.py

% from https://tableaufriction.blogspot.com/2012/11/finally-you-can-use-tableau-data-colors.html  
\definecolor{tableau_blue}{RGB}{31,119,180}
\definecolor{tableau_orange}{RGB}{255,127,14}
\definecolor{tableau_red}{RGB}{214,39,40}
\definecolor{tableau_grey}{RGB}{127,127,127}
\definecolor{tableau_green}{RGB}{44,160,44}
\definecolor{tableau_purple}{RGB}{148,103,189}
\definecolor{tableau_cyan}{RGB}{23,190,207}

% vertical lines right and left of hack (not used)
\newcommand{\hackstartline}{
\draw [tableau_red, very thick, dashed] 
(axis cs:\hackstart,\pgfkeysvalueof{/pgfplots/ymin}) -- 
(axis cs:\hackstart,\pgfkeysvalueof{/pgfplots/ymax});}

\newcommand{\hackendline}{
\draw [tableau_red, very thick, dashed] 
(axis cs:\hackend,\pgfkeysvalueof{/pgfplots/ymin}) -- 
(axis cs:\hackend,\pgfkeysvalueof{/pgfplots/ymax});}

\newcommand{\hacklines}{\hackstartline\hackendline}

% shaded area over hack
\newcommand{\hackshade}{
\draw [fill=tableau_red, opacity=.06] 
(axis cs:\hackstart,\pgfkeysvalueof{/pgfplots/ymin}) -- 
(axis cs:\hackend,\pgfkeysvalueof{/pgfplots/ymin}) -- 
(axis cs:\hackend,\pgfkeysvalueof{/pgfplots/ymax}) -- 
(axis cs:\hackstart,\pgfkeysvalueof{/pgfplots/ymax});
}

\begin{groupplot}[group style={
        group name=my plots,
        group size=1 by 4,
        xlabels at=edge bottom,
        xticklabels at=edge bottom,
        vertical sep=15pt,
    },
    xlabel={Time (s)},
    legend style={nodes={scale=0.6, transform shape}},
    width=\linewidth,
    height=3.3cm,
    enlarge x limits=false,
    tick pos=left,
    xticklabel style={font=\scriptsize},
    yticklabel style={font=\scriptsize,rotate=90,anchor=base,yshift=0.2cm,scaled y ticks=false, /pgf/number format/fixed},
    minor y tick num=0,
    title style={yshift=-8pt}
]
%\nextgroupplot[ylabel=voltage,legend pos=north east, ymax=1.06, ymin=0.96,try min ticks=3]\hackshade
\nextgroupplot[ylabel=voltage (p.u.),legend pos=south west, ymax=1.05, ymin=0.92, ytick={0.93, 0.98, 1.04},legend columns=1]\hackshade
\addplot [smooth, tableau_blue, no markers] table [x=, y=va, col sep=comma] {\evalhist};
\addplot [smooth, tableau_green, no markers] table [x=, y=vb, col sep=comma] {\evalhist};
\addplot [smooth, tableau_cyan, no markers] table [x=, y=vc, col sep=comma] {\evalhist};
\addlegendentry{Phase A}
\addlegendentry{Phase B}
\addlegendentry{Phase C}
\coordinate (hackstart) at (axis cs:\hackstart,\pgfkeysvalueof{/pgfplots/ymax});
\coordinate (hackend) at (axis cs:\hackend,\pgfkeysvalueof{/pgfplots/ymax});

%\hacklines
%\addlegendentry{voltage}

%\nextgroupplot[ylabel=VU,legend pos=north east, ymin=-0.005, ymax=0.06, ytick={0, 0.02, 0.04}]\hackshade
\nextgroupplot[ylabel=VU (\%),legend pos=north east, ymin=0, ymax=6, ytick={0, 2, 4, 6}, title=Largest Network Voltage Imbalance]\hackshade
\addplot [tableau_red, no markers] table [x=,  y expr=\thisrow{u_worst}*100, col sep=comma] {\evalhist};
%\hacklines
%\addlegendentry{oscillation observer}

\nextgroupplot[ylabel=$\Delta \eta$, legend pos=north east, legend columns=4,try min ticks=1,ymin=-0.11,ymax=0.11, title=Agent action]\hackshade
\addplot [tableau_blue, no markers] table [x=, y=ta_0, col sep=comma] {\evalhist};
\addplot [tableau_green, no markers] table [x=, y=tb_0, col sep=comma] {\evalhist};
\addplot [tableau_cyan, no markers] table [x=, y=tc_0, col sep=comma] {\evalhist};
\addlegendentry{Offset phase A}
\addlegendentry{Offset phase B}
\addlegendentry{Offset phase C}

% \nextgroupplot[ylabel=tap position, legend pos=north east, legend columns=4,try min ticks=1, ymin=-4,ymax=17, title=Voltage Regulator]\hackshade
% \addplot [tableau_purple, no markers] table [x=, y=reg, col sep=comma] {\evalhist};
%\addplot [tableau_red, no markers] table [x=, y=ht_all, col sep=comma] {\evalhist};
%\hacklines

% \nextgroupplot[ylabel=reward, legend pos=south east, legend columns=1, ymax=5, ymin=-50]\hackshade
% \addplot [tableau_purple, no markers] table [x=, y=component_y, col sep=comma] {\evalhist};
% \addplot [tableau_green, no markers] table [x=, y=component_oa, col sep=comma] {\evalhist};
% \addplot [tableau_orange, no markers] table [x=, y=component_init, col sep=comma] {\evalhist};
% \addplot [tableau_red, no markers] table [x=, y=component_pset_pmax, col sep=comma] {\evalhist};
% \addplot [tableau_blue, no markers] table [x=, y=total_reward, col sep=comma] {\evalhist};

% %\hacklines
% \addlegendentry{Oscillation}
% \addlegendentry{Last action}
% \addlegendentry{Initial action}
% \addlegendentry{P curtailment}
% \addlegendentry{Total reward}

% \nextgroupplot[ylabel=kVar, legend pos=south east, legend columns=4,ytick={-1000, 0, 1000}, ymin=-2000,ymax=1500, title= Total Reactive Power Injection]\hackshade
\nextgroupplot[ylabel=MVar, legend pos=south east, legend columns=4, ytick={-1, 0, 1}, ymin=-2,ymax=1.5, title= Total DER Reactive Power Injection]\hackshade
\addplot [tableau_blue, no markers] table [x=, y expr=-\thisrow{qouta_total}/1000, col sep=comma] {\evalhist};
\addplot [tableau_red, no markers] table [x=, y expr=-\thisrow{qoutb_total}/1000, col sep=comma] {\evalhist};
\addplot [tableau_green, no markers] table [x=, y expr=-\thisrow{qoutc_total}/1000, col sep=comma] {\evalhist};

%\addplot [name path=upper,draw=none] table[x=,y expr=-\thisrow{qouta_mean}+\thisrow{qouta_std}] {\evalhist};
%\addplot [name path=lower,draw=none] table[x=,y expr=-\thisrow{qouta_mean}-\thisrow{qouta_std}] {\evalhist};
%\addplot [fill=blue!10] fill between[of=upper and lower];
%\errorband[tableau_blue, opacity=0.5]{\evalhist}{}{qouta_mean}{qouta_std};
%\errorband[tableau_blue, opacity=0.4]{\evalhist}{}{mean}{std}

\addlegendentry{Phase A}
\addlegendentry{Phase B}
\addlegendentry{Phase C}

\end{groupplot}

\draw[latex-latex, thick, tableau_red, yshift=0.5cm] (hackstart) to node [auto] {attack} (hackend);

\end{tikzpicture}
\caption{30\% DER imbalance attack with ADAM-based ARS}
\label{fig:ars_vu}
\end{figure}
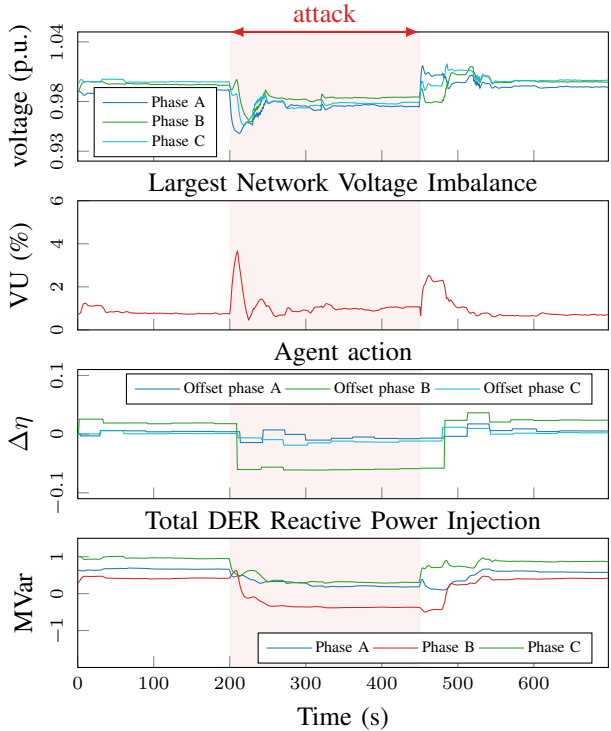

While a linear policy is sufficient to mitigate oscillation attacks, mitigation of imbalance attacks requires a more complex policy structure (i.e., a neural network). An attack on 30\% of DER VV/VW controllers aimed at creating large voltage imbalances is depicted in Fig. \ref{fig:baseline_vu}.  This baseline case demonstrates the effect of the attack in the system without utilizing the policy trained by Adam-ARS to control the remaining DER smart inverters in the system.  Fig. \ref{fig:ars_vu} shows the effect of a neural network-based policy trained by Adam-ARS. As is shown in the figure, the policy significantly reduces the largest voltage imbalance from 3\% to less than 1\%.  Hyperparameters for this experiment are shown in Table \ref{table:hyperparams2}.

\section{Conclusions}
\label{sec:conclusions}

This paper explored the use of an Adam-based Augmented Random Search (ARS) algorithm to directly search for optimal policies that manage DER smart inverter Volt-VAR/Volt-Watt control functions to mitigate the effects of cyber attacks.  In the event that a malicious entity gains the ability to manipulate the VV/VW settings in a portion of DER in a given system, the trained optimal policy was shown to be effective in reconfiguring the remaining \emph{non-compromised} DER VV/VW controllers in the system to mitigate large oscillations and large imbalances in system voltages.  Compared to previous attempts, we demonstrated that the Adam-ARS approach can learn optimal policies significantly faster than Proximal Policy Optimization (PPO), faster than ARS, and with less variance in the reward compared to ARS.  Additionally, the Adam-ARS algorithm is able to learn a \emph{linear} policy for defense against voltage oscillation attacks.  The results herein indicate that training optimal control policies for DER cybersecurity can be performed with fewer computational resources than previously believed, thereby allowing grid-managing entities with less expertise and financial resources the ability to use this technology for their systems.

While in this work we have demonstrated the effectiveness in adapting the learning rate during the training process, future work will explore adapting the variance of the exploration noise used for finite-difference approximation of the reward gradient.  We believe that this feature may further reduce training time and possibly yield superior rewards compared to the proposed algorithm.

\bibliographystyle{IEEEtran}
\bibliography{references}

\begin{thebibliography}{10}
\providecommand{\url}[1]{#1}
\csname url@samestyle\endcsname
\providecommand{\newblock}{\relax}
\providecommand{\bibinfo}[2]{#2}
\providecommand{\BIBentrySTDinterwordspacing}{\spaceskip=0pt\relax}
\providecommand{\BIBentryALTinterwordstretchfactor}{4}
\providecommand{\BIBentryALTinterwordspacing}{\spaceskip=\fontdimen2\font plus
\BIBentryALTinterwordstretchfactor\fontdimen3\font minus
  \fontdimen4\font\relax}
\providecommand{\BIBforeignlanguage}[2]{{%
\expandafter\ifx\csname l@#1\endcsname\relax
\typeout{** WARNING: IEEEtran.bst: No hyphenation pattern has been}%
\typeout{** loaded for the language `#1'. Using the pattern for}%
\typeout{** the default language instead.}%
\else
\language=\csname l@#1\endcsname
\fi
#2}}
\providecommand{\BIBdecl}{\relax}
\BIBdecl

\bibitem{npr_clean_energy}
``{Proposed Clean Energy Standard Could End Power Plant Greenhouse Gas
  Emissions By 2035},''
  \url{https://www.npr.org/2021/08/11/1026831067/proposed-clean-energy-standard-could-end-power-plant-greenhouse-gas-emissions-by},
  [Online; published Aug. 11, 2021].

\bibitem{doe_cut_solar}
``{DOE Announces Goal to Cut Solar Costs by More than Half by 2030},''
  \url{https://www.energy.gov/articles/doe-announces-goal-cut-solar-costs-more-half-2030},
  [Online; published Mar. 25, 2021].

\bibitem{IEEE_1547}
{IEEE Standards Coordinating Committee 21}, ``{IEEE Standard for
  Interconnection and Interoperability of Distributed Energy Resources with
  Associated Electric Power Systems Interfaces},'' \emph{IEEE Std 1547-2018
  (Revision of IEEE Std 1547-2003)}, pp. 1--138, April 2018.

\bibitem{sahoo2019cyber}
S.~Sahoo, T.~Dragi{\v{c}}evi{\'c}, and F.~Blaabjerg, ``{Cyber Security in
  Control of Grid-Tied Power Electronic Converters--Challenges and
  Vulnerabilities},'' \emph{IEEE Journal of Emerging and Selected Topics in
  Power Electronics}, 2019.

\bibitem{spectrum2015}
``{800,000 Microinverters Remotely Retrofitted on Oahu in One Day},''
  \url{https://spectrum.ieee.org/energywise/green-tech/solar/in-one-day-800000-microinverters-remotely-retrofitted-on-oahu},
  [Online; accessed June-2019].

\bibitem{roberts2020deep}
C.~Roberts, S.-T. Ngo, A.~Milesi, S.~Peisert, D.~Arnold, S.~Saha, A.~Scaglione,
  N.~Johnson, A.~Kocheturov, and D.~Fradkin, ``{Deep Reinforcement Learning for
  DER Cyber-Attack Mitigation},'' in \emph{2020 IEEE International Conference
  on Communications, Control, and Computing Technologies for Smart Grids
  (SmartGridComm)}.\hskip 1em plus 0.5em minus 0.4em\relax IEEE, 2020, pp.
  1--7.

\bibitem{roberts2021deep_unbalance}
C.~Roberts, S.-T. Ngo, A.~Milesi, A.~Scaglione, S.~Peisert, and D.~Arnold,
  ``{Deep Reinforcement Learning for Mitigating Cyber-Physical DER Voltage
  Unbalance Attacks},'' in \emph{2021 American Control Conference (ACC)}.\hskip
  1em plus 0.5em minus 0.4em\relax IEEE, 2021, pp. 2861--2867.

\bibitem{spall2003stochastic}
J.~C. Spall, \emph{Introduction to Stochastic Search and Optimization}.\hskip
  1em plus 0.5em minus 0.4em\relax John Wiley \& Sons, Ltd, 2003.

\bibitem{krstic2003extremum}
M.~Krstic, \emph{Real‐Time Optimization by Extremum‐Seeking Control}.\hskip
  1em plus 0.5em minus 0.4em\relax John Wiley \& Sons, Ltd, 2003.

\bibitem{salimans2017evolution}
T.~Salimans, J.~Ho, X.~Chen, S.~Sidor, and I.~Sutskever, ``{Evolution
  Strategies as a Scalable Alternative to Reinforcement Learning},''
  \url{https://arxiv.org/abs/1703.03864}, 2017.

\bibitem{mania2018simple}
H.~Mania, A.~Guy, and B.~Recht, ``Simple random search provides a competitive
  approach to reinforcement learning,'' \url{https://arxiv.org/abs/1803.07055},
  2018.

\bibitem{duchi2011adaptive}
J.~Duchi, E.~Hazan, and Y.~Singer, ``Adaptive subgradient methods for online
  learning and stochastic optimization,'' \emph{Journal of Machine Learning
  Research}, vol.~12, no.~61, pp. 2121--2159, 2011.

\bibitem{inverter2016}
B.~Seal, ``{Common Functions for Smart Inverters, 4th Ed.}'' Electric Power
  Research Institute, Tech. Rep. 3002008217, 2017.

\bibitem{kingma2017adam}
D.~P. Kingma and J.~Ba, ``Adam: A method for stochastic optimization,''
  \url{https://arxiv.org/abs/1412.6980}, 2017.

\bibitem{hintonRMS}
G.~Hinton, N.~Srivastava, and K.~Swersky, ``{Lecture notes on Neural Networks
  for Machine Learning - Lecture 6a}.''

\bibitem{pmlr-v9-glorot10a}
X.~Glorot and Y.~Bengio, ``Understanding the difficulty of training deep
  feedforward neural networks,'' in \emph{Proceedings of the Thirteenth
  International Conference on Artificial Intelligence and Statistics}, ser.
  Proceedings of Machine Learning Research, Y.~W. Teh and M.~Titterington,
  Eds., vol.~9.\hskip 1em plus 0.5em minus 0.4em\relax Chia Laguna Resort,
  Sardinia, Italy: PMLR, 13--15 May 2010, pp. 249--256.

\end{thebibliography}

\appendix

\begin{table}[ht!]
\centering
\begin{tabular}{|llll|}
\hline
\textbf{Hyperparameter} & \textbf{ARS} & \textbf{ADAM-based ARS} & \textbf{PPO}\\
\hline

$\alpha$ & $1\times 10^{-2}$ & $5\times 10^{-2}$ & $1\times 10^{-3}$\\
$\mu$ & $3\times 10^{-2}$ & $3\times 10^{-2}$ & -\\ 
$\beta_0$ & $0.9$ & 0.9 & -\\ 
$\beta_1$ & 0.999 & 0.999 & -\\ 
$\lambda_{ADAM}$ & $1\times 10^{-8}$ & $1\times 10^{-8}$ & -\\ 
$\gamma$ & --  & --  & 0.5 \\
$\lambda_{PPO}$ & -- & -- & 0.95\\
$\epsilon$ & -- & -- & 0.1 \\
episodes & 16 & 16 & 8\\ 
activation function & tanh & tanh & tanh\\
hidden layers & dense(16,16) & dense(16,16) & dense(16,16) \\ 
$\sigma_y$ & 300 & 300 & 300\\
$\sigma_u$ & 300 & 300 & 300\\
$\sigma_a$ & 0.5 & 0.5 & 0.5\\
$\sigma_0$ & 1 & 1 & 1\\
$\sigma_p$ & 1 & 1 & 1\\ 

\hline

\end{tabular}
\vspace{0.1cm}
\caption{Hyperparameters of the network, training and reward for unbalance attack}
\label{table:hyperparams1}
\end{table}

\begin{table}[ht!]
\centering
\begin{tabular}{|llll|}
\hline
\textbf{Hyperparameter} & \textbf{ARS} & \textbf{ADAM-based ARS} & \textbf{PPO}\\
\hline

$\alpha$ & $3\times 10^{-3}$ & $3\times 10^{-3}$  & $1\times 10^{-3}$\\
$\mu$ & $1\times 10^{-2}$ & $1\times 10^{-2}$ & -\\ 
$\beta_0$ & $0.9$ & 0.9 & -\\ 
$\beta_1$ & 0.999 & 0.999 & -\\ 
$\lambda_{ADAM}$ & $1\times 10^{-8}$ & $1\times 10^{-8}$ & -\\
$\gamma$ & --  & --  & 0.5 \\
$\lambda_{PPO}$ & -- & -- & 0.95\\
$\epsilon$ & -- & -- & 0.1 \\
episodes & 8 & 8 & 8\\ 
activation function & -- & -- & tanh\\
hidden layers & linear & linear & dense (16, 16) \\ 
$\sigma_y$ & 300 & 300 & 300\\
$\sigma_u$ & 300 & 300 & 300\\
$\sigma_a$ & 0.5 & 0.5 & 0.5\\
$\sigma_0$ & 1 & 1 & 1\\
$\sigma_p$ & 1 & 1 & 1\\ 

\hline

\end{tabular}
\vspace{0.1cm}
\caption{Hyperparameters of the network, training and reward for oscillation attack}
\label{table:hyperparams2}
\end{table}

\end{document}